# Towards a Computed Collateral Circulation Score in Ischemic Stroke

Marta Kersten-Oertel[1], Ali Alamer[2], Vladimir Fonov[1], Benjamin W.Y. Lo[3], Donatella Tampieri[2,3], D. Louis Collins[1,3]

[1]McConnell Brain Imaging Centre, Montreal Neurological Institute & Hospital (MNI/H)
[2]Deptartment of Diagnostic and Interventional Neuroradiology, MNI/H
[3]Department of Neurology and Neurosurgery, McGill University, MNI/H

**Abstract.** Stroke is the second leading cause of disability worldwide. In order to minimize disability, the goal of stroke treatment is to preserve tissue in the area where blood supply is decreased but sufficient to stave off cell death. Thrombectomy has been shown to offer fast and efficient reperfusion with high recanalization rates. However, due to the inherent risks of thrombectomy, indications including evidence of good collateral circulation should be present. Currently, methods for evaluating collateral circulation are limited. In this work, we present an automated technique to compute a collateral circulation score based on differences seen in mean intensities between left and right cerebral hemispheres in 4D angiography images. In this preliminary study, we analyzed scans of 28 subjects: 4 with normal collateral circulation, 7 with good, 10 with intermediate, and 7 with poor collateral circulation, (based on radiologist score) and found a good correlation between the computed score and radiologist score ($r^2$ = 0.71) and good separation between good and intermediate/poor groups. This work shows promise as the first step in building an automated collateral circulation score.

## 1   Introduction

According to the World Heart Federation, each year over 15 million people suffer from stroke, with 6 million dying as a result, and 5 million becoming permanently disabled[1]. The two main types of stroke are: (1) hemorrhagic, due to bleeding, and (2) ischemic, due to a lack of blood flow. In this paper, we focus on ischemic stroke, which accounts for approximately 80% of all stroke cases. In ischemic stroke, where poor blood flow to the brain causes neuronal cell death, the goal of treatment is to restore blood flow to preserve tissue in the ischemic *penumbra*, where blood flow is decreased but sufficient enough to stave off infarction (i.e. cell death).

It has been shown that recanalization, i.e. restoring blood flow, is the most important modifiable prognostic predictor for a favourable outcome in ischemic stroke [1]. Timely restoration of regional blood flow can help salvage threatened tissue, reducing cell death and ultimately patient disabilities. If recanalization is successful, the chance of a favorable outcome is increased 4-fold compared to patients without recanalization. Furthermore, there is a 4-fold decrease in mortality among patients with successful recanalization [2]. Strategies for recanalization include the use of thrombolytic agents such as IV drugs (e.g. tissue plasminogen activator (tPA)) and/or mechanical techniques such as distal or proximal *thrombectomies* and stent retrievers. A *thrombectomy* is an intervention where a long catheter with a mechanical device attached to the tip, is used to remove a clot. The short procedure times, high recanalization rates, and the possibility of fast efficient blood flow restoration, make the use of thrombectomy attractive. However, the inherent risks associated with thrombectomy must be considered and only patients with certain indications, including a large penumbra and small infarct and good collateral circulation should undergo such interventions. In other words, in order to achieve high rates of favorable outcomes for mechanical recanalization interventions, advanced selection of patients based on pre-treatment imaging is crucial [3-6].

Collateral circulation (i.e. *collaterals)* is defined as a supplementary vascular network that is dynamically recruited when there is an arterial occlusion (e.g. a clot). Over the last several years, numerous studies (e.g. [7], [8], [9]) have stressed the importance of collateral evaluation prior to interventions. Although the penumbra and infarct region are easily identifiable on clinical maps created from angiography scans, including measurements of cerebral blood flow and volume (CBF and CBV), mean transit time (MTT) and time to peak (TTP); the best way to evaluate collateral circulation is still debated. Currently, collaterals are typically visually evaluated on angiography scans (X-ray, CTA, and MRA), however, there is no consensus on which type of imaging modality should be used [10] and there are several different grading scores with no current gold standard [3]. For assessment on CTA, the collateral score is based on visual inspection of the images by a radiologist and can be graded using scoring systems by the American Society of Interventional and Therapeutic Neuroradiology/Society of Interventional Radiology (ASITN/SIR), the Alberta Stroke Program Early CT Score (ASPECTS), Christoforidis *et al.* and Miteff *et al*. [3]. The score is purely subjective, and typically classified as: good, intermediate and poor, based on the presence of any delay of opacification of the peripheral arteries in the territory of the stroke, as recommended by the ESCAPE trial [4]. Due to the fact that collateral circulation recruitment is crucial in influencing outcome in stroke, a robust and accurate collateral circulation metric is needed. In this paper, we present a novel automatic image processing method to compute a collateral score based on dynamic 4D CTA images.

## 2 Materials and Methods

### 2.1 Study Design and Patients

We selected 28 patients who had undergone imaging according to the scanning protocol for stroke (described below) at the Montreal Neurological Institute and Hospital (MNI/H). Our study included four control subjects (*normal*), and 24 patients that had suffered a stroke. The collateral circulation score of the patients was consensually agreed on by two radiologist's as being *good*, *intermediate* or *poor* using ASPECTS[2]. The collateral circulation is scored as follows: a score of *good* is given for 100% collateral supply of the occluded MCA territory; *intermediate* score is given when collateral supply fills more than 50% but less then 100% of the occluded MCA territory; a *poor* score indicates collateral supply that fills less then 50% but more than 0% of the occluded MCA territory (**Fig. 1**). Among patients in this preliminary study, we had 7 good, 10 intermediate, and 7 poor collateral scores.

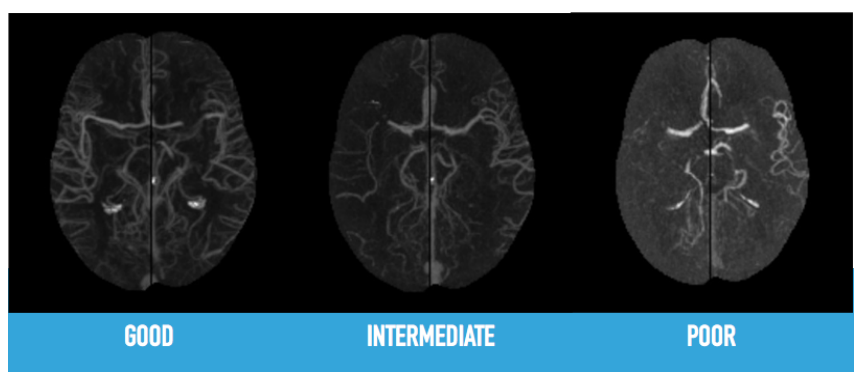

**Fig. 1.** Example maximum intensity projections (MIPs) of different collateral circulation scores. The collateral circulation is scored as *good* (100% collateral supply of the occluded MCA territory); *intermediate* (collateral supply filling >50% but <100% of the occluded MCA territory) or *poor* (collateral supply filling ≤50% but >0% of the occluded MCA territory).

### 2.2 Scanning Protocol

In all patients, computed tomography (CT) imaging was performed on Toshiba's Aquilion ONE 320-row detector 640-slice cone beam CT (Toshiba medical systems, Tokyo, Japan) that is able to acquire isotropic volumes of the entire brain with a single rotation of the gantry. The routine stroke protocol performs a series of intermittent volume scans over a period of 60 seconds with a scanning speed of 0.75 s/rotation. This protocol provides whole brain perfusion and whole brain dynamic vascular analysis in one examination. A total of 19 volumes are acquired, where the first is acquired prior to injection and is used as a mask for the dynamic subtraction. Next, a series of low-dose scans are performed, first for every two seconds during the arterial phase, and then spaced out to every 5 seconds to capture the slower venous phase.

Isovue-370 (Iopamidol) was used as non-ionic and low osmolar contrast medium (Iodine content, 370 mg/ml). The injection protocol is 50 ml contrast medium at rate of 5 ml/s followed by second phase of 50 ml saline at similar rate using a power injector. The injector is started together with the scan to ensure correct timing of the scan sequence. For evaluation of the collateral circulation, all volumes are reconstructed into temporal transverse maximum intensity projection (MIPs) images with slice thickness of 20 mm, which provides time-resolved images of the arterial, parenchymal and venous phases. As a result, the extent and velocity of the collateral filling can be evaluated with dynamic-CTA as opposite to single phase-CTA.

---

[2] http://www.aspectsinstroke.com/

### 2.3 Image Processing & Analysis

We developed a custom pipeline, shown in (**Fig. 2**), using tools from the MINC toolkit[3] to automatically process the 4D-CTA data.

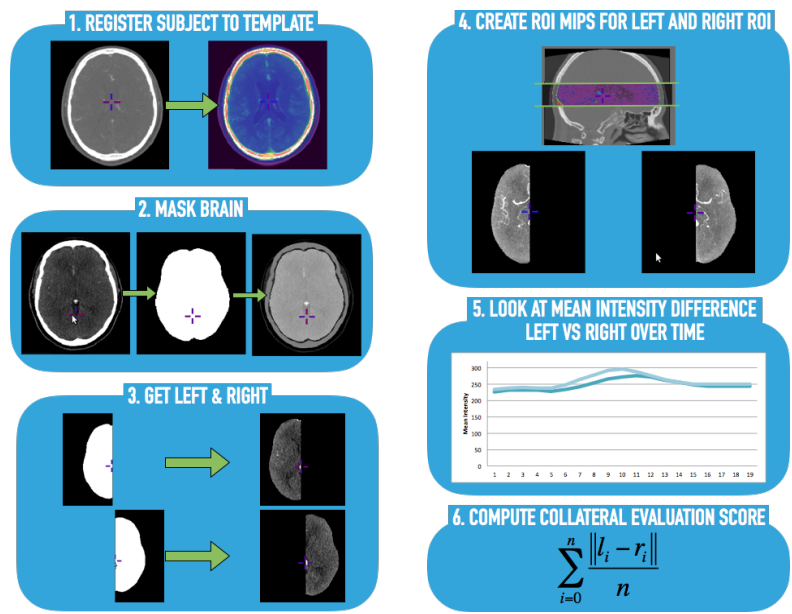

**Fig. 2.** The pipeline for computing the collateral circulation score involves registering each CTA volume to the population-specific template and looking at the left versus right hemisphere's mean intensity over time.

First, a population-specific non-linear average template was created from 12 normal subjects that had undergone the same dynamic 4D-CTA imaging protocol (**Fig. 3**. The template was created using the algorithm described in [11]. The template resolution was $1 \times 1 \times 1$ mm$^3$. Each of the 28 subjects' dynamic volumes (18 per subject) were then registered to the template. This was done by (1) computing a non-linear registration of the first volume to the template and (2) computing a rigid linear registration between each of the remaining volumes to the first volume (to account for any patient movement during the acquisition). These transformations were concatenated and used to accurately map every dynamic volume to the template space. Registra-tion, which is driven mostly by the bright signal (i.e. bone and major blood vessels), was successful and accurate for all subjects. We used the obtained transformation to warp the brain mask to identify the subject's brain and separate it into left and right hemispheres, and to define a region of interest (20 axial slices from approximately below the nose to the brow). A transverse MIP of these regions is created, and the mean intensity of the MIP is computed separately for the left and right hemispheres. The difference between left and right mean intensities for each volume is summed and divided by the number of volumes in the sequence as follows:

$$\frac{\sum_{i=1}^{n} \| l_i - r_i \|}{n}$$

where $l_i$ is the mean intensity of the left MIP and $r_i$ is the mean intensity of the right MIP and $n$ is the number of volumes.

---

[3] http://bic-mni.github.io/

When the difference in means between left and right is low, there is good symmetry and therefore good collaterals, when the difference is high there are not as many bright vessels in one hemisphere. Therefore, our computed score relates the numbers of vessels seen in the left and right hemispheres over time.

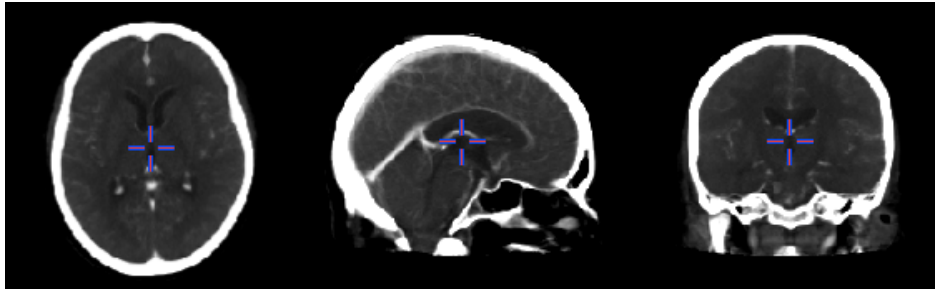

**Fig. 3.** Average non-linear template created from 12 normal subjects.

## 3    Results

We analyzed our data using an analysis of variance (ANOVA) test and found a good correlation (*r-square* = 0.71) between radiologist score (categorical data: normal (0), good (1), intermediate (2), poor (3)) and the numerical computed collateral score (**Fig. 4**). Furthermore, post-hoc Tukey-Kramer tests showed that there was significant difference between normal and poor collaterals ($p < 0.0001$), and normal and intermediate collaterals ($p = 0.0003$), as well as between good and intermediate ($p = 0.0004$) and good and poor collaterals ($p < 0.0001$). As expected there was no significant difference between normal and good collaterals ($p = 0.8352$). For good and normal groups, the computed collateral score ranged from 0.2 to 1.0 (based on the radiologist scoring). Where a low score indicates a good symmetry between left and right circulation and either no ischemic stroke (i.e. normal controls) or very good collateral circulation (radiologist rating of good). There was no significant difference between inter-mediate and poor collateral scores ($p = 0.5037$). Ideally, there would also be a good separation between intermediate and poor scores. Yet, the intermediate and poor scores based on the radiologist evaluations spanned between 1.1 and 2.1, with a mean of 1.45 (SD = 0.38) for intermediate and 1.68 (SD = 0.3) for poor. The means and standard deviations for the group scores are shown in **Table 1**.

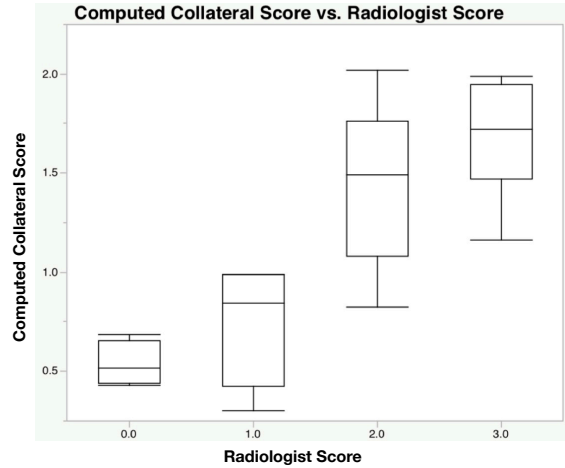

**Fig. 4.** Average plotted scores, *x-axis* is the radiologist score (RD Score), *y-axis* is the score computed using our pipeline. Normal refers to our control subjects and and good, intermediate and poor to stroke patients with radiologist evaluated collateral scores. There is a good correlation between the computed score and radiologist score (*r-square* = 0.71). Furthermore, there was a good separation between good (*score = 0*), and intermediate (*score = 2*) and poor group scores (*score = 3*).

| Collateral Score | Number | Mean | Std Dev | Std Error |
|---|---|---|---|---|
| normal | 4 | 0.54 | 0.11 | 0.06 |
| good | 7 | 0.70 | 0.28 | 0.11 |
| intermediate | 10 | 1.45 | 0.39 | 0.12 |
| poor | 6 | 1.68 | 0.30 | 0.12 |

**Table 1.** The mean and standard deviation and standard errors for each of the collateral score groups.

On closer inspection of intermediate and poor subjects, we found that in several cases which would be misclassified (according to radiologist rating), there was often other visible elements of the brain anatomy that influenced the score. For example, calcifications in the pituitary gland, ventricles or the falx that were not symmetric but appear bright on CTA (similar contrast to that of bone), affected the mean intensity and consecutively the computed score, to be lower or higher depending on whether they were on the stroke-affected side of the brain or not. Furthermore, sinuses that appear on one side also affected the scores. Given the accurate registration results (which were visually verified) we believe that this is the major shortcoming of the method and what is driving the differences between the radiologist score and computed score.

In **Figs. 5** and **6** we show examples of the MIPs over the 4D CTA volumes, the mean intensity values for left versus right hemispheres and the computed and radiologist score. We further point out some calcifications and other asymmetries that could affect the collateral score. As can be seen, the graphs give a good visualization of the differences between left and right hemispheres; the further apart the left and right mean curves, the poorer the collateral score.

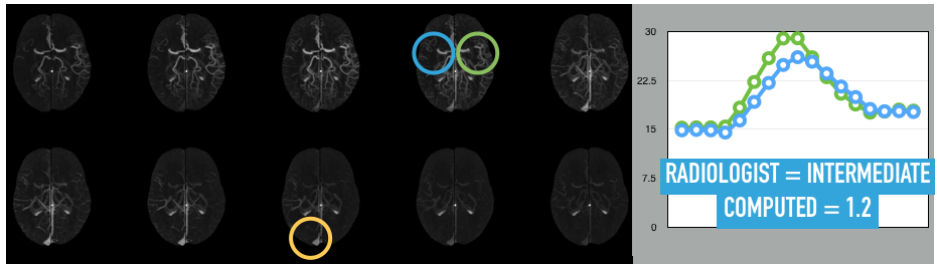

**Fig. 5.** MIPs of each of the 3D CTA volumes over time (right) the mean intensity plot of right (green) versus left (blue) hemisphere and the computed and radiologist score. The sinus which is more visible on the left, the stroke affected side, (circled in orange) slightly affects the score to be slightly higher than it might otherwise be. Note only 10 of the 18 volumes are shown.

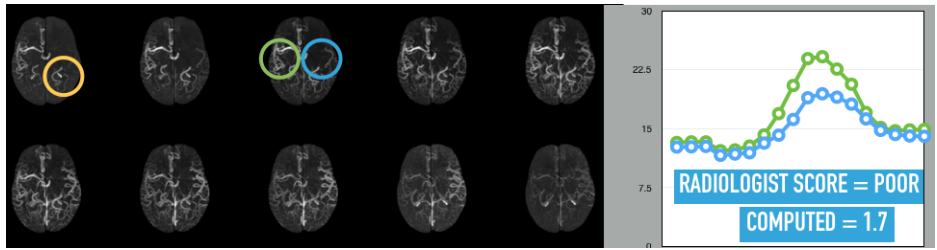

**Fig. 6.** MIPs of each of the 3D CTA volumes over time (right) the mean intensity plot of right (green) vs. left (blue) hemisphere and the computed radiologist score. In the yellow circle see a calcification that is bigger on the right then the left and therefore maybe artificially lower the difference between hemispheres. Note due to space only 10 of 18 volumes are shown.

## 4   Conclusions and Future Work

Although a good correlation between radiologist and computed score was found, and good separation between good and poor scores, future work will further refine the computed score by looking not only at the differences between left and right hemispheres but also at the collaterals solely in the MCA territory. This will remove many of the bright voxels that should not be taken into consideration in the collateral circulation score, and should further help to separate intermediate from poor collaterals. This is necessary as in clinical practice, it is more important to differentiate between good and intermediate collaterals versus poor collateral circulation because it has been shown that the results of thrombectomy are worse in those with poor collaterals [4].

Upon examination of the MIPs created from our pipeline, as well as the visual graphs, in a few cases the radiologists changed their score from intermediate to poor or vice-versa. This suggests that there would be some inter-rater disagreement about scores. In future work, we will evaluate the inter-rater and intra-rater variability in labelling collaterals. Furthermore, we will explore machine learning and classification techniques that should be well suited to this problem.

In this work we have developed a simple automatic pipeline to compute a collateral circulation score. We believe this is a first step in developing a robust score that can be used by radiologists and neurosurgeons to determine the best course of treatment and to predict patient outcomes.